\documentclass[a4paper]{article}
\RequirePackage[english]{babel}
\RequirePackage[latin1]{inputenc}
\RequirePackage[T1]{fontenc}
\RequirePackage{amsbsy}
\RequirePackage{amssymb}
\RequirePackage{amsmath}
\RequirePackage{mathrsfs}
\RequirePackage{bm}
\newtheorem{theorem}{Theorem}
\begin{document}
\title{\huge{\bf{On the problem of Unicity in\\ Einstein-Sciama-Kibble\\ Theory}}}
\author{Luca Fabbri\\ 
\footnotesize (Theory Group, INFN \& Department of Physics, Bologna, ITALY)}
\date{}
\maketitle
\begin{abstract}
We consider the ESK theory, based on the principle for which the space is filled with matter fields in such a way that Cartan torsion is spin; in the geometry in which Cartan torsion tensor is completely antisymmetric, spin has to be completely antisymmetric: we will show how the complete antisymmetry of spin constraints the form of possible matter field theories as to allow the special case of the simplest one solely, all other most general cases being excluded.
\end{abstract}
\section*{Introduction}
Although geometry has been developed as to be the most general formalism describing quantities belonging to a given space, so that one could be led to think that it talks about no particular ones, it is indeed structured in such a way that a couple of special tensors, the Riemann tensor and the Cartan torsion tensor, do emerge. 

In these spaces, it is possible to define physical matter fields, and according to the principle put forward by Einstein first and by Sciama and Kibble later, for which the matter content of a space influences the structure of the space itself, it is possible to think that from physical matter fields we can construct physical tensors to be coupled to the geometrical Riemann and Cartan tensors, where this coupling is defined by least-order derivative field equations. Although this theory limits the form a matter field theory can have, it does not say anything about its actual content, leaving the issue of which physical theory is the true description of the material universe widely open; nevertheless, if we consider the prescription according to which energy is curvature and spin is torsion, and the case in which torsion is completely antisymmetric, then spin has to be completely antisymmetric as well.

In this paper we will see how the complete antisymmetry of the spin constraints the form matter field theories are allowed to have.
\section{Unicity in Einstein-Sciama-Kibble Theory}
In the geometry of tensors, the procedure of lowering and raising indices in tensors is allowed by the introduction of the two fundamental tensors $g_{\alpha\beta}$ and $g^{\alpha\beta}$ such that $g_{\alpha\beta}=g_{\beta\alpha}$ and $g^{\alpha\beta}=g^{\beta\alpha}$ and also $g_{\alpha\beta}g^{\beta\mu}=\delta_{\alpha}^{\mu}$ where $\delta^{\mu}_{\nu}$ is the Kronecker Unity tensor used to rename indices in tensors. The tensors $g$ happen to have all the features that characterize metric properties, and so they are called Metric Tensors.

The covariant differential structures $D_{\mu}$ are defined after the introduction of the connections $\Gamma^{\alpha}_{\mu\nu}$; the procedure of lowering and raising indices in tensors that are derivatives of some other tensors is possible in general whenever it is given the condition $D_{\mu}g_{\alpha\beta}=0$, and since this condition does not dependent on the specific derivative, it is justified to require it to hold for any derivative, and so for any connection. Because this condition is supposed to hold for any connection, then the connections are called Metric Connections.

If we consider that all the connections are metric, then it is easy to prove that their Cartan torsion tensor $Q_{\alpha\mu\rho}$ has to be completely antisymmetric; see for example \cite{f} for a more extensive discussion.

Further, given the fundamental Levi-Civita metric connection $\Lambda^{\alpha}_{\mu\nu}=\Lambda^{\alpha}_{\nu\mu}$ written in terms of the metric alone, then we have that
\begin{eqnarray}
\Gamma^{\alpha}_{\mu\nu}=\Lambda^{\alpha}_{\mu\nu}+\frac{1}{2}Q^{\alpha}_{\phantom{\alpha}\mu\nu}
\label{conn}
\end{eqnarray}
is the decomposition of the most general connection, and we will write the covariant derivative calculated with respect to the Levi-Civita metric connection $\Lambda^{\alpha}_{\mu\nu}$ as $\nabla_{\mu}$, keeping the covariant derivative with respect to the connection $\Gamma^{\alpha}_{\mu\nu}$ as $D_{\mu}$; also, given the fundamental Riemann curvature tensor $R^{\alpha}_{\phantom{\alpha}\rho\alpha\sigma}$ written in terms of the Levi-Civita metric connection alone, we have that 
\begin{eqnarray}
\nonumber
G^{\mu}_{\phantom{\mu} \kappa \rho \beta}=R^{\mu}_{\phantom{\mu} \kappa \rho \beta}+\frac{1}{2}(\nabla_{\rho}Q^{\mu}_{\phantom{\mu} \kappa \beta}
-\nabla_{\beta}Q^{\mu}_{\phantom{\mu} \kappa \rho})+\\
+\frac{1}{4}(Q^{\mu}_{\phantom{\mu} \sigma \rho}Q^{\sigma}_{\phantom{\sigma} \kappa \beta}
-Q^{\mu}_{\phantom{\mu}\sigma \beta}Q^{\sigma}_{\phantom{\sigma} \kappa \rho})
\end{eqnarray}
is the decomposition of the most general Riemann tensor written in terms of the most general connection in this case.

Because of its symmetry properties, both Riemann curvature tensor and Riemann tensor admit only one independent contraction given by Ricci curvature tensor $R^{\alpha}_{\phantom{\alpha}\rho\alpha\sigma}=R_{\rho\sigma}$ and Ricci tensor $G^{\alpha}_{\phantom{\alpha}\rho\alpha\sigma}=G_{\rho\sigma}$, which admit one contraction given by Ricci curvature scalar $R_{\rho\sigma}g^{\rho\sigma}=R$ and Ricci scalar $G_{\rho\sigma}g^{\rho\sigma}=G$, decomposable as 
\begin{eqnarray}
G_{\kappa\beta}=R_{\kappa\beta}
+\frac{1}{2}\nabla_{\mu}Q^{\mu}_{\phantom{\mu} \kappa \beta}
-\frac{1}{4}Q^{\mu\sigma}_{\phantom{\mu\sigma}\beta}Q_{\mu\sigma\kappa}
\end{eqnarray}
and also as 
\begin{eqnarray}
G=R-\frac{1}{4}Q^{\mu\sigma\beta}Q_{\mu\sigma\beta}
\end{eqnarray}
which will be useful in the following; again, these decompositions are the most general for completely antisymmetric Cartan torsion tensor.

Riemann and Ricci tensors together with the completely antisymmetric Cartan torsion tensor are such that they verify the geometrical identities
\begin{eqnarray}
\nonumber
&(D_{\kappa}Q^{\rho}_{\phantom{\rho}\mu \nu}
+Q^{\rho}_{\phantom{\rho}\kappa \pi}Q^{\pi}_{\phantom{\pi}\mu \nu}
+G^{\rho}_{\phantom{\rho}\kappa\mu\nu})
+(D_{\nu}Q^{\rho}_{\phantom{\rho} \kappa \mu}
+Q^{\rho}_{\phantom{\rho}\nu \pi}Q^{\pi}_{\phantom{\pi}\kappa \mu}
+G^{\rho}_{\phantom{\rho}\nu\kappa\mu})+\\
&+(D_{\mu}Q^{\rho}_{\phantom{\rho} \nu \kappa}
+Q^{\rho}_{\phantom{\rho}\mu \pi}Q^{\pi}_{\phantom{\pi} \nu \kappa}
+G^{\rho}_{\phantom{\rho}\mu\nu\kappa})\equiv0
\label{torsion}
\end{eqnarray}
and
\begin{eqnarray}
\nonumber
&(D_{\mu}G^{\nu}_{\phantom{\nu}\iota \kappa \rho}
-G^{\nu}_{\phantom{\nu}\iota \beta \mu}Q^{\beta}_{\phantom{\beta}\kappa\rho})
+(D_{\kappa}G^{\nu}_{\phantom{\nu}\iota \rho \mu}
-G^{\nu}_{\phantom{\nu}\iota \beta \kappa}Q^{\beta}_{\phantom{\beta}\rho\mu})+\\
&+(D_{\rho}G^{\nu}_{\phantom{\nu}\iota \mu \kappa}
-G^{\nu}_{\phantom{\nu}\iota \beta \rho}Q^{\beta}_{\phantom{\beta}\mu\kappa})\equiv0
\label{curvatorsiontorsion}
\end{eqnarray}
called Jacobi-Bianchi identities; finally
\begin{eqnarray}
D_{\rho}Q^{\rho\mu\nu}+(G^{\nu\mu}-\frac{1}{2}g^{\nu\mu}G)-(G^{\mu\nu}-\frac{1}{2}g^{\mu\nu}G)\equiv0
\label{torsionfc}
\end{eqnarray}
and
\begin{eqnarray}
D_{\rho}(G^{\rho\kappa}-\frac{1}{2}g^{\rho\kappa}G)
+(G_{\rho\beta}-\frac{1}{2}g_{\rho\beta}G)Q^{\rho\beta\kappa}
+\frac{1}{2}Q_{\nu\rho\beta}G^{\nu\rho\beta\kappa}\equiv0
\label{curvatorsiontorsionfc}
\end{eqnarray}
are the fully contracted Jacobi-Bianchi identities.

The fundamental principle we are going to assume will be that the structure of a space is determined by its matter content in minimal interaction, in the sense that the fundamental geometrical quantities have a dynamics whose source is given by physical fields as to verify the energy-curvature and spin-torsion coupling; according to this prescription 
\begin{eqnarray}
G^{\sigma\rho}-\frac{1}{2}g^{\sigma\rho}G=-8 \pi K \ T^{\sigma\rho}
\label{e}
\end{eqnarray}
and
\begin{eqnarray}
Q^{\sigma\rho\theta}=16\pi K \ S^{\sigma\rho\theta}
\label{sk}
\end{eqnarray}
where $K$ is the gravitational constant, and where $T^{\sigma\rho}$ is the Energy-Momentum Tensor and $S^{\sigma\rho\theta}$ is the Spin Tensor, and they are thus called Einstein-Sciama-Kibble Field Equations, defining the Einstein-Sciama-Kibble theories. 

Note that due to the decomposition above, we can write the first equation in terms of the second equation as
\begin{eqnarray}
\nonumber
R^{\sigma\rho}-\frac{1}{2}g^{\sigma\rho}R=\\
=-8\pi K\left[8\pi K\left(\frac{1}{2}g^{\sigma\rho}S_{\mu\nu\beta}S^{\mu\nu\beta}-
S_{\mu\nu}^{\phantom{\mu\nu}\sigma}S^{\mu\nu\rho}\right)+\nabla_{\mu}S^{\mu\sigma\rho}+T^{\sigma\rho}\right]
\end{eqnarray}
in which we see how these field equations represent a modification of Einstein field equations for gravity, because they are essentially given by the equations
\begin{eqnarray}
R^{\sigma\rho}-\frac{1}{2}g^{\sigma\rho}R=-8 \pi K \ T^{\sigma\rho}
\end{eqnarray}
which are the Einstein Field Equations, defining the Einstein theory, once a spin interaction is added.

The Einstein theory realizes a principle of geometrization for which matter fields are coupled to geometrical quantities, and in this special case energy is coupled to curvature; but we know that matter fields have spin beside energy, whereas geometry has torsion beside curvature, and thus the same principle of geometrization of Einstein theory is generalized in the Sciama-Kibble extension: thus, if we want to maintain this idea of geometrization in the most general situation, we are led to the Einstein-Sciama-Kibble theory necessarily.

\subsection{Unicity of Completely Antisymmetric Spins}
The fact that the geometrical background allows the existence of both energy-momentum and spin tensors permits Fermionic matter field theories to be defined; see \cite{d} and \cite{l} for general considerations about completely antisymmetric spin.

\subsubsection{Unicity of Fermionic Matter Field Theories}
In the geometrical background outlined, and with this principle, since the content of matter is thought to be represented by Fermionic fields, defined in terms of spinor fields transforming according to spinorial representations of the Lorentz group, the general procedure to write down Fermionic matter field theories is to consider spinors, and to build from them the energy-momentum tensor $T^{\mu\nu}$ and the spin tensor $S^{\rho\mu\nu}$ such that it is completely antisymmetric, that is $S^{\rho\mu\nu}=\frac{1}{6}S^{[\rho\mu\nu]}$ where the squared parentheses denote the antisymmetrization or commutation of those indices, and such that the conservation laws
\begin{eqnarray}
\nabla_{\rho}S^{\rho\mu\nu}+\frac{1}{2}T^{[\mu\nu]}=0
\label{t-spin}
\end{eqnarray}
and
\begin{eqnarray}
D_{\sigma}T^{\sigma\rho}
-T_{\beta\sigma}Q^{\sigma\beta\rho}-S_{\beta\mu\sigma}G^{\mu\sigma\beta\rho}=0
\label{mt-energy}
\end{eqnarray}
are verified.

As before, we note that when the spin vanishes the energy-momentum tensor $T^{\mu\nu}$ must be symmetric, that is $T^{\mu\nu}=\frac{1}{2}T^{\{\mu\nu\}}$ where the curled parentheses denote the symmetrization or anticommutation of those indices, and such that it is conserved
\begin{eqnarray}
\nabla_{\sigma}T^{\sigma\rho}=0
\end{eqnarray}
identically, as the standard theory requires.

In general, since these conservation laws are not geometrical identities, they will result in dynamical constraints that will be satisfied only if the spinor fields themselves verify a given set of dynamical constraints called spinor field equations, or Fermionic Matter Field Equations.

Obviously, \emph{a priori} there is the possibility to have many possible Fermionic matter field theories defined by their fundamental Fermionic matter field equations.

\paragraph{Examples of higher-order derivative field equations.} Higher-Order dynamical field theories will be characterized by equations for the fields that are allowed to have derivatives to be of order higher than the least one. For simplicity, we will study only the first order after the least one, that is the second order; second order field equations have then two derivatives, and hence two indices that have to be contracted.

The simplest contractions we can have are when the two indices are contracted with the constant $\sigma^{\rho\alpha}$ matrices that are the spinorial representation of the generators of the Lorentz group, or alternatively, when they are contracted with each other, and accordingly we are going to consider these two cases.

\subparagraph{Second-order fields with external contraction.} This theory is defined by second-order field equations in which the second-order derivative term is contracted with the constant $\sigma^{\rho\alpha}$ matrices.

Defining the conjugate as $\overline{\psi}\equiv \psi^{\dagger}C$ where $C$ is a constant matrix such that $C^{\dagger}=C$ and $C^{-1}\sigma_{\rho\alpha}^{\dagger}C=-\sigma_{\rho\alpha}$, we can define the energy-momentum and spin tensors as
\begin{eqnarray}
\nonumber
T^{\alpha\mu}=D^{\mu}\overline{\psi}\sigma^{\alpha\rho}D_{\rho}\psi
+D_{\rho}\overline{\psi}\sigma^{\rho\alpha}D^{\mu}\psi-\\
-g^{\alpha\mu}\left(D_{\rho}\overline{\psi}\sigma^{\rho\sigma}D_{\sigma}\psi-m^{2}\psi^{2}\right)
\label{antisymmetricdenergy}
\end{eqnarray}
and
\begin{eqnarray}
S^{\alpha\rho\beta}=\frac{1}{2}
\left(D_{\mu}\overline{\psi}\sigma^{\mu\alpha}\sigma^{\rho\beta}\psi
-\overline{\psi}\sigma^{\rho\beta}\sigma^{\alpha\mu}D_{\mu}\psi\right)
\label{antisymmetricdspin}
\end{eqnarray}
non-completely antisymmetric; field equations will be
\begin{eqnarray}
\sigma^{\mu\nu}D_{\mu}D_{\nu}\psi=-m^{2}\psi
\label{antisymmetricdequation}
\end{eqnarray}
where $m$ is the mass of the field, and we have that the conservation laws (\ref{t-spin}) and (\ref{mt-energy}) are verified.

The expression for the spin is not completely antisymmetric, and it can be completely antisymmetrized by the condition
\begin{eqnarray}
D_{\mu}\overline{\psi}(\sigma^{\mu\alpha}\sigma^{\rho\beta}
+\sigma^{\mu\rho}\sigma^{\alpha\beta})\psi
-\overline{\psi}(\sigma^{\rho\beta}\sigma^{\alpha\mu}
+\sigma^{\alpha\beta}\sigma^{\rho\mu})D_{\mu}\psi=0
\label{antisymmetricdconstraint}
\end{eqnarray}
which gives rise to the algebraic solution
\begin{eqnarray}
\sigma^{\mu\alpha}\sigma^{\rho\beta}+\sigma^{\mu\rho}\sigma^{\alpha\beta}=0
\label{antisymmetricdconstraintalgebraic}
\end{eqnarray}
between $\sigma$ matrices, and this relationship implies that $\sigma^{\mu\alpha}\sigma_{\mu\beta}=0$, so that it gives $[\sigma^{\mu\alpha},\sigma_{\mu\rho}]=0$ and finally $\sigma^{\alpha\beta}=0$ which is a trivial condition.

\subparagraph{Second-order fields with internal contraction.} This theory is defined by second-order field equation in which the second-order derivative term is contracted with itself.

Defining the conjugate as in the previous case, we can define the energy-momentum and spin tensors to be
\begin{eqnarray}
T^{\alpha\mu}=D^{\mu}\overline{\psi}D^{\alpha}\psi+D^{\alpha}\overline{\psi}D^{\mu}\psi
-g^{\alpha\mu}\left(D_{\rho}\overline{\psi}D^{\rho}\psi-m^{2}\psi^{2}\right)
\label{symmetricdenergy}
\end{eqnarray}
symmetric and
\begin{eqnarray}
S^{\mu\rho\beta}=\frac{1}{2}
\left(D^{\mu}\overline{\psi}\sigma^{\rho\beta}\psi
-\overline{\psi}\sigma^{\rho\beta}D^{\mu}\psi\right)
\label{symmetricdspin}
\end{eqnarray}
non-completely antisymmetric; field equations will then be
\begin{eqnarray}
D^{2}\psi=-m^{2}\psi
\label{symmetricdequation}
\end{eqnarray}
where $m$ is the mass of the field, and we have that the conservation laws (\ref{t-spin}) and (\ref{mt-energy}) are verified.

Even if the energy-momentum tensor is symmetric, and consequently the spin tensor is truly covariantly conserved, nonetheless the spin is not completely antisymmetric, although it can be antisymmetrized through the condition
\begin{eqnarray}
D^{\rho}\overline{\psi}\sigma^{\mu\beta}\psi
-\overline{\psi}\sigma^{\mu\beta}D^{\rho}\psi+
D^{\mu}\overline{\psi}\sigma^{\rho\beta}\psi
-\overline{\psi}\sigma^{\rho\beta}D^{\mu}\psi=0
\label{symmetricdconstraint}
\end{eqnarray}
so that, taking into account this additional condition, the spin turns out to be completely antisymmetric, but always such that with the same symmetric energy-momentum tensor, and the same spinor field equations, we have that the conservation laws (\ref{t-spin}) and (\ref{mt-energy}) are verified.

However, from this additional condition it is not possible to extract an algebraic solution between the $\sigma$s, and so this will turn out to be an additional constraint on the fields themselves.

So, it is clear that higher-order derivative field equations are not well defined, and in consequence we will turn to the least-order derivative field equations.

\paragraph{Examples of least-order derivative field equations.} Least-Order dynamical field theories are characterized by equations for the fields that have derivatives of the least order possible, that is they will be first-order theories, and hence one index will have to be contracted.

For these theories, then, a set of constant $\gamma^{\mu}$ matrices must be introduced in order for the unique index of the only derivative to be contracted away; for the moment we will make no assumption about these matrices.

The fundamental fields can be classified according to their transformation law, and we will consider the cases in which the transformation laws will be given by $\psi'^{a}=S\Lambda^{a}_{b}\psi^{b}$, $\psi'=S\psi S^{-1}$ and $\psi'=S\psi$, respectively.

\subparagraph{Type $\boldsymbol{\psi'^{a}=S\Lambda^{a}_{b}\psi^{b}}$ field.} In this theory, the fundamental fields are defined to transform as $\psi'^{a}=S\Lambda^{a}_{b}\psi^{b}$ where $S$ is the spinorial representation of the Lorentz group and $\Lambda$ is the transformation of the Lorentz group, and so it has one spinorial and one tensorial index; we will call them Dirac-Rarita-Schwinger Spinors.

Defining the conjugate as $\overline{\psi_{a}}\equiv \psi^{\dagger}_{a}C$, we can define the energy-momentum and spin tensors as
\begin{eqnarray}
T^{\sigma\rho}=
\frac{i}{2}\left(\overline{\psi^{\alpha}}\gamma^{\sigma}D^{\rho}\psi_{\alpha}
-D^{\rho}\overline{\psi^{\alpha}}\gamma^{\sigma}\psi_{\alpha}\right)
\label{drsenergy}
\end{eqnarray}
and
\begin{eqnarray}
S^{\alpha\theta\mu}=\frac{i}{2}[(\overline{\psi^{\theta}}\gamma^{\alpha}\psi^{\mu}
-\overline{\psi^{\mu}}\gamma^{\alpha}\psi^{\theta})
+\frac{1}{2}\overline{\psi^{\sigma}}\{\gamma^{\alpha},\sigma^{\theta\mu}\}\psi_{\sigma}]
\label{drsspin}
\end{eqnarray}
non-completely antisymmetric; Dirac-Rarita-Schwinger spinor field equations will be
\begin{eqnarray}
i\gamma^{\mu}D_{\mu}\psi^{\alpha}=m\psi^{\alpha}
\label{drsequation}
\end{eqnarray}
where $m$ is the mass of the field, and we have that the conservation laws for the spin and the energy-momentum are verified.

However, the previous form of the spin tensor (\ref{drsspin}) is not completely antisymmetric, but it can be antisymmetrized developing the constraints
\begin{eqnarray}
\{\gamma^{\mu},\sigma^{\sigma\beta}\}+\{\gamma^{\sigma},\sigma^{\mu\beta}\}=0
\label{drscondition}
\end{eqnarray}
and
\begin{eqnarray}
\overline{\psi^{\theta}}\gamma^{\alpha}\psi^{\mu}
-\overline{\psi^{\mu}}\gamma^{\alpha}\psi^{\theta}
+\overline{\psi^{\theta}}\gamma^{\mu}\psi^{\alpha}
-\overline{\psi^{\alpha}}\gamma^{\mu}\psi^{\theta}=0
\label{drsconstraint}
\end{eqnarray}
so that, taking into account these additional conditions, the spin turns out to be completely antisymmetric, and such that with the same energy-momentum tensor, and the same Dirac-Rarita-Schwinger spinor field equations, we have that the conservation laws are verified.

Of the two conditions above, the second turns out to be an algebraic relationship between dynamical fields and, since it is specifically imposed upon fields in order to reduce their general dynamical properties, it appears to be an unphysical constraint.

We note that alternative types of Dirac-Rarita-Schwinger spinor field theories can be defined, as exemplified in \cite{d-z}, and what is interesting is that the unphysical constraint $(\ref{drsconstraint})$ is developed in each version of Dirac-Rarita-Schwinger spinor field theory.

So in all the cases discussed here above, there are unphysical constraints that are developed, making the Dirac-Rarita-Schwinger Fermionic field theories in their most general forms unphysical matter field theories.

\subparagraph{Type $\boldsymbol{\psi'=S\psi S^{-1}}$ field.} In this theory, the fundamental fields are defined to transform as $\psi'=S\psi S^{-1}$ where $S$ is the spinorial representation of the Lorentz group, and so it has two spinorial indices; we will call them Dirac-Dirac Spinors.

Defining the conjugate as $\overline{\psi}\equiv C^{-1}\psi^{\dagger}C$, we can define the energy-momentum and spin tensors as
\begin{eqnarray}
T^{\sigma\rho}=
\frac{i}{2}\mathrm{Tr}\left(\overline{\psi}\gamma^{\sigma}D^{\rho}\psi
-D^{\rho}\overline{\psi}\gamma^{\sigma}\psi\right)
\label{2denergy}
\end{eqnarray}
and
\begin{eqnarray}
S^{\mu\sigma\beta}=\frac{i}{2}[-\mathrm{Tr}(\overline{\psi}\gamma^{\mu}\psi\sigma^{\sigma\beta})+
\frac{1}{2}\mathrm{Tr}(\overline{\psi}\{\gamma^{\mu},\sigma^{\sigma\beta}\}\psi)]
\label{2dspin}
\end{eqnarray}
non-completely antisymmetric; Dirac-Dirac spinor field equations will be
\begin{eqnarray}
i\gamma^{\mu}D_{\mu}\psi=m\psi
\label{2dequation}
\end{eqnarray}
with contraction
\begin{eqnarray}
i\mathrm{Tr}(\gamma^{\mu}D_{\mu}\psi)=m\mathrm{Tr}\psi
\label{2dequationcontracted}
\end{eqnarray}
where $m$ is the mass of the field, and we see that the conservation laws for the spin and the energy-momentum are verified.

Although the previous form of the spin tensor (\ref{2dspin}) is not completely antisymmetric, it can be antisymmetrized developing the constraints
\begin{eqnarray}
\{\gamma^{\mu},\sigma^{\sigma\beta}\}+\{\gamma^{\sigma},\sigma^{\mu\beta}\}=0
\label{2dcondition}
\end{eqnarray}
and
\begin{eqnarray}
\mathrm{Tr}(\overline{\psi}\gamma^{\alpha}\psi\sigma^{\mu\rho}+
\overline{\psi}\gamma^{\mu}\psi\sigma^{\alpha\rho})=0
\label{2dconstraint}
\end{eqnarray}
so that, taking into account these additional conditions, the spin turns out to be completely antisymmetric, but always such that with the same energy-momentum tensor, and the same Dirac-Dirac spinor field equations, we have that the conservation laws are verified.

Again, of these two additional conditions, the second is an algebraic relationship between dynamical fields, and so another unphysical constraint, making the Dirac-Dirac Fermionic field theory considered in its most general form an unphysical matter field theory, as for the previous case.

\subparagraph{Type $\boldsymbol{\psi'=S\psi}$ field.} In this theory, the fundamental fields are defined to transform as $\psi'=S\psi$ where $S$ is the spinorial representation of the Lorentz group, and so it has one spinorial index; we will call them Dirac Spinors.

Defining the conjugate as $\overline{\psi}\equiv \psi^{\dagger}C$, we can define the energy-momentum and spin tensors as
\begin{eqnarray}
T^{\sigma\rho}=
\frac{i}{2}\left(\overline{\psi}\gamma^{\sigma}D^{\rho}\psi
-D^{\rho}\overline{\psi}\gamma^{\sigma}\psi\right)
\label{denergy}
\end{eqnarray}
and
\begin{eqnarray}
S^{\mu\sigma\beta}=
\frac{i}{4}\overline{\psi}\{\gamma^{\mu},\sigma^{\sigma\beta}\}\psi
\label{dspin}
\end{eqnarray}
which is not completely antisymmetric yet; Dirac spinor field equations are
\begin{eqnarray}
i\gamma^{\mu}D_{\mu}\psi=m\psi
\label{dequation}
\end{eqnarray}
where $m$ is the mass of the field, and we see that the conservation laws are verified.

We see that the condition of antisymmetry for any couple of indices of the spin tensor can now turn into the condition
\begin{eqnarray}
\{\gamma^{\mu},\sigma^{\sigma\beta}\}+\{\gamma^{\sigma},\sigma^{\mu\beta}\}=0
\label{condition}
\end{eqnarray}
which is an algebraic constraint upon constant matrices; this condition not only has none of the problems we discussed above, but it is a condition that can determine the relationship between the $\gamma$ and the $\sigma$ matrices that we have not required yet.

To do this, we have to take into account that this last condition has to be improved by the condition of constancy of the $\gamma$ matrices, and by the fundamental relationships that define the $\sigma$ matrices to be the spinorial representation of the generators of the Lorentz group; the question now is whether we can use all these relationships to get some new information about the constant matrices involved.

We have then the following
\begin{theorem}
If $\{\sigma^{\sigma\beta}\}$ is the set of constant matrices that are the spinorial representation of the generators of the Lorentz group, and it is introduced $\{\gamma^{\alpha}\}$ as a set of constant matrices such that the condition  
\begin{eqnarray}
\{\gamma^{\mu},\sigma^{\sigma\beta}\}+\{\gamma^{\sigma},\sigma^{\mu\beta}\}=0
\end{eqnarray}
holds, then we have the following relationships
\begin{eqnarray}
A\sigma^{\alpha\beta}=\sigma^{\alpha\beta}A=\frac{1}{4}[\gamma^{\alpha},\gamma^{\beta}]
\end{eqnarray}
with the conditions
\begin{eqnarray}
\{\gamma^{\alpha},\gamma^{\beta}\}=2Ag^{\alpha\beta}
\end{eqnarray}
and
\begin{eqnarray}
[\gamma^{\mu},A]=0
\end{eqnarray}
in terms of the constant matrix $A$ still unknown.\\
$\square$
\footnotesize
\textit{Proof.} Considering the hypothesis of the theorem, we have that $\sigma^{\sigma\beta}$ are the constant matrices that are the spinorial representation of the generators of the Lorentz group, and $\gamma^{\alpha}$ are constant matrices for which the condition  
\begin{eqnarray}
\{\sigma^{\sigma\beta},\gamma^{\mu}\}+\{\sigma^{\mu\beta},\gamma^{\sigma}\}=0
\label{ah}
\end{eqnarray}
holds; being them constant matrices, they have also to verify the constancy conditions
\begin{eqnarray}
[\sigma^{\sigma\beta},\gamma^{\mu}]=g^{\mu\beta}\gamma^{\sigma}-g^{\mu\sigma}\gamma^{\beta}
\label{ac}
\end{eqnarray}
which combined together with the hypothesis gives
\begin{eqnarray}
\nonumber
2(\sigma^{\mu\nu}\gamma^{\beta}+\sigma^{\beta\nu}\gamma^{\mu})=g^{\mu\nu}\gamma^{\beta}+g^{\beta\nu}\gamma^{\mu}-2g^{\mu\beta}\gamma^{\nu}
\end{eqnarray}
whose contraction is
\begin{eqnarray}
\sigma^{\mu\nu}\gamma_{\nu}=\frac{(n-1)}{2}\gamma^{\mu}
\end{eqnarray}
and this formula will help us in simplifying or introducing the $\sigma$ matrices whenever useful.

From this last formula, we get that
\begin{eqnarray}
\nonumber
&\frac{(n-1)}{2}[\gamma^{\beta},\gamma^{\alpha}]=[\sigma^{\beta\nu}\gamma_{\nu},\gamma^{\alpha}]=
\sigma^{\beta\nu}[\gamma_{\nu},\gamma^{\alpha}]+[\sigma^{\beta\nu},\gamma^{\alpha}]\gamma_{\nu}=\\
\nonumber
&=\sigma^{\beta\nu}[\gamma_{\nu},\gamma^{\alpha}]+\gamma^{\beta}\gamma^{\alpha}
-g^{\alpha\beta}\gamma^{\nu}\gamma_{\nu}
\end{eqnarray}
and the relationship
\begin{eqnarray}
\nonumber
\frac{(n-1)}{2}[\gamma^{\beta},\gamma^{\alpha}]=
\sigma^{\beta\nu}[\gamma_{\nu},\gamma^{\alpha}]+\gamma^{\beta}\gamma^{\alpha}
-g^{\alpha\beta}\gamma^{\nu}\gamma_{\nu}
\end{eqnarray}
can be decomposed into its symmetric and antisymmetric part, where the latter one gives
\begin{eqnarray}
\nonumber
(n-2)[\gamma^{\beta},\gamma^{\alpha}]=
\sigma^{\beta\nu}[\gamma_{\nu},\gamma^{\alpha}]
-\sigma^{\alpha\nu}[\gamma_{\nu},\gamma^{\beta}]
\end{eqnarray}
and so
\begin{eqnarray}
\nonumber
&(n-2)[\gamma^{\beta},\gamma^{\alpha}]=
\sigma^{\beta\nu}\gamma_{\nu}\gamma^{\alpha}
-\sigma^{\beta\nu}\gamma^{\alpha}\gamma_{\nu}
-\sigma^{\alpha\nu}\gamma_{\nu}\gamma^{\beta}
\sigma^{\alpha\nu}\gamma^{\beta}\gamma_{\nu}=\\
\nonumber
&=\frac{(n-1)}{2}[\gamma^{\beta},\gamma^{\alpha}]
-\{\sigma^{\beta\nu},\gamma^{\alpha}\}\gamma_{\nu}
+[\sigma^{\alpha\nu},\gamma^{\beta}]\gamma_{\nu}
+\gamma^{\beta}\sigma^{\alpha\nu}\gamma_{\nu}
+\gamma^{\alpha}\sigma^{\beta\nu}\gamma_{\nu}
\end{eqnarray}
identically. By using again the hypothesis we get
\begin{eqnarray}
\nonumber
&\frac{(n-3)}{2}[\gamma^{\beta},\gamma^{\alpha}]=
-\{\sigma^{\beta\nu},\gamma^{\alpha}\}\gamma_{\nu}
+[\sigma^{\alpha\nu},\gamma^{\beta}]\gamma_{\nu}
+\gamma^{\beta}\sigma^{\alpha\nu}\gamma_{\nu}
+\gamma^{\alpha}\sigma^{\beta\nu}\gamma_{\nu}=\\
\nonumber
&=\{\sigma^{\beta\alpha},\gamma^{\nu}\}\gamma_{\nu}
+\gamma^{\alpha}\gamma^{\beta}-g^{\alpha\beta}\gamma^{\nu}\gamma_{\nu}
+\frac{(n-1)}{2}\{\gamma^{\beta},\gamma^{\alpha}\}
\end{eqnarray}
whose symmetric part reads
\begin{eqnarray}
\nonumber
0=-2g^{\alpha\beta}\gamma^{\nu}\gamma_{\nu}
+n\{\gamma^{\beta},\gamma^{\alpha}\}
\end{eqnarray}
that is
\begin{eqnarray}
\nonumber
n\{\gamma^{\alpha},\gamma^{\beta}\}=2g^{\alpha\beta}\gamma^{\nu}\gamma_{\nu}
\end{eqnarray}
irreducible, and which can be cast into the form
\begin{eqnarray}
\{\gamma^{\alpha},\gamma^{\beta}\}=2g^{\alpha\beta}A
\label{acondition}
\end{eqnarray}
by calling $\gamma^{\nu}\gamma_{\nu}=nA$ for any $n$-dimensional space.

By using this condition (\ref{acondition}) and the previous relationships, we can now see that
\begin{eqnarray}
\nonumber
&2\sigma^{\beta}_{\phantom{\beta}\nu}A=\sigma^{\mu\nu}g_{\beta\mu}A2=
\sigma^{\mu\nu}\{\gamma_{\mu},\gamma_{\beta}\}=
\sigma^{\mu\nu}\gamma_{\mu}\gamma_{\beta}+\sigma^{\mu\nu}\gamma_{\beta}\gamma_{\mu}=\\
\nonumber
&=-\frac{(n-1)}{2}\gamma^{\nu}\gamma_{\beta}+[\sigma^{\mu\nu},\gamma_{\beta}]\gamma_{\mu}+\gamma^{\beta}\sigma^{\mu\nu}\gamma_{\mu}=\\
\nonumber
&=-\frac{(n-1)}{2}\{\gamma^{\nu},\gamma_{\beta}\}+
\delta^{\nu}_{\beta}\gamma^{\mu}\gamma_{\mu}-\gamma^{\nu}\gamma_{\beta}=\\
\nonumber
&=\frac{1}{2}[\gamma^{\beta},\gamma_{\nu}]
\end{eqnarray}
so that 
\begin{eqnarray}
\sigma^{\beta\nu}A=\frac{1}{4}[\gamma^{\beta},\gamma^{\nu}]
\end{eqnarray}
and an analogous path would have given the relationship
\begin{eqnarray}
A\sigma^{\beta\nu}=\frac{1}{4}[\gamma^{\beta},\gamma^{\nu}]
\end{eqnarray}
as well; this gives in particular the fact that the matrix $A$ commutes with all the $\sigma$s. Finally, by plugging $A$ into the hypothesis, we find immediately that $A$ also commutes with all the $\gamma$s, and the theorem is proved.

This proof works for any representation, in any dimension.
\normalsize
$\blacksquare$
\end{theorem}
and so the matrix $A$ is not the identity, but a more general one.

Thus we see that although transforming according to the same transformation law, and governed by dynamical field equations that are formally analogous, what we called Dirac spinors are not really the usual Dirac spinor fields, but one of their generalizations.

In this case, it is with no other constraint that the spin (\ref{dspin}) is completely antisymmetric, while field equations are still verified, and thus Dirac Fermionic field theory is perfectly allowed.

From all these examples, we have shown that there have to be no tensorial indices, and no more than one spinorial index, so that we will be left with a field having one spinorial index only, that is we have the least-spin field; and from all these examples, we have shown that these fields have to satisfy no higher-order field equations, that is the dynamics of these least-spin matter fields is governed by least-order derivative field equations: so we are left with the least-spin matter least-order dynamical field theory as the only possible Fermionic matter field theory.

This Fermionic matter field theory is the one characterized by the complete antisymmetry of spin achieved in the most general case, while both for spin and energy the conservation laws are fulfilled as well.

Thus the simplest Fermionic matter field theory alone is allowed, all the others are forbidden in general.
\section*{Conclusion}
In this paper, we have shown that most of the Fermionic theories are excluded, since higher-order derivative higher-spins have in general a non-completely antisymmetric spin tensor, as it should be in order to correspond to the completely antisymmetric Cartan torsion tensor; hence, since the only theory of matter whose spin is antisymmetric is the least-spin least-order dynamical Fermionic matter field theory, then Dirac theory is the only one that is allowed for the completely antisymmetric Cartan torsion tensor in the scheme of the Einstein-Sciama-Kibble theory.

And so, whereas the existence of Fermionic field theories is permitted by the introduction of the spin, their restriction to the simplest one is permitted by the complete antisymmetry of the spin itself.

\end{document}